\documentstyle[prl,aps,psfig,epsfig,multicol]{revtex}
\begin{document}
\def\be{\begin{equation}}
\def\ee{\end{equation}}
\def\bc{\begin{center}}
\def\ec{\end{center}}
\def\bea{\begin{eqnarray}}
\def\eea{\end{eqnarray}}
\newcommand{\bleq}{\ifpreprintsty
                   \else
                   \end{multicols}\vspace*{-3.5ex}\widetext{\tiny
                   \noindent\begin{tabular}[t]{c|}
                   \parbox{0.493\hsize}{~} \\ \hline \end{tabular}}
                   \fi}
\newcommand{\eleq}{\ifpreprintsty
                   \else
                   {\tiny\hspace*{\fill}\begin{tabular}[t]{|c}\hline
                    \parbox{0.49\hsize}{~} \\
                    \end{tabular}}\vspace*{-2.5ex}\begin{multicols}{2}
                    \narrowtext \fi}

\title{Growing Cayley trees described by Fermi distribution}
\author{Ginestra Bianconi}
\address{Department of Physics, University of Notre Dame, Notre
Dame,Indiana
 46556,USA }
\maketitle

\begin{abstract}
We introduce a model for   growing   Cayley trees with thermal
noise. The evolution of these hierarchical networks reduces  to
the Eden model and the invasion percolation model in the  limit
$T\rightarrow 0$,   $T\rightarrow \infty$ respectively. We show
that the distribution of the bond strengths (energies) is
described by the  Fermi statistics. We discuss the relation of the
present results with
 the scale-free networks described by Bose statistics.
\end{abstract}
\begin{multicols}{2}
\narrowtext

\section{Introduction}

Recently it has been shown that  Bose statistics\cite{PRL} can be
used to describe a scale-free network\cite{BA,RMP} with fitness of
the nodes\cite{Fitness}. Since scale-free networks are
continuously growing and develop a power-law connectivity
distribution it is interesting to investigate their relation with
self-organized processes\cite{IP,SOC,BS}. In order to address this
problem we present a model of invasion percolation with
temperature defined on a Cayley tree, self-organized in the limit
$T\rightarrow 0$. Our results show that the model can be solved
analytically using the same technique used in the case of the
scale-free networks described by Bose statistics.

The invasion percolation model\cite{IP} is the most famous and
simple example of evolution with quenched disorder. It describes
the displacement of a fluid in a porous medium.
  The porous medium is given by
a random network constituted by bonds with different strengths $p$
chosen with uniform probability in the interval $(0,1)$.
 The classical asymptotic structure generated in this way
is a fractal  and the distribution of the strength values at the
interface  converges in time to a step function $\theta(p-p_c)$,
where $p_c$ is the percolation threshold of the static percolation
problem.  In order to include the effect of fluctuations on the
dynamics of invasion, present in a real stochastic cases, we
 include a temperature-like
noise $T$\cite{Vergeles1,Vergeles2,Caldarelli,Gabrielli}.
 A  structure in which the
invasion percolation dynamics can be defined is a
Cayley-tree\cite{IPCT,Vanderwalle}, also known as the Bethe
lattice. In this structure there are no loops, and the number of
nodes in the bulk are of the same order of magnitude as the nodes
at the interface. Therefore   a Cayley tree  is considered to be a
good representation of a  $d=\infty$ space and it is used in mean
field calculations $\cite{Thorpe}$ and in the study of branching
processes $\cite{Harris}$.
 In this
work we find that the distribution of bond strengths at the
interface is no more a step function but it is described by the
Fermi distribution  with temperature $T$ where the bond strength
plays the role of  energy. By comparison with scale-free networks
following the Bose statistics I show that both networks grow
continuously in time: in the power-law network at each time a node
is connected to the network by $m$ links while in the Cayley tree
model at each time a node grows giving rise to $m$ new nodes. The
dynamics of the two networks change in time awarding the fitter
nodes in the power-law network or choosing the less fit nodes to
grow. At the same time in the power-law network the distribution
of the energies of the chosen nodes converges to  a Bose
distribution while in a Cayley tree model the distribution of the
energies at the interface converges to a Fermi distribution.

\begin{figure}
\centerline{\epsfxsize=2.5in \epsfbox{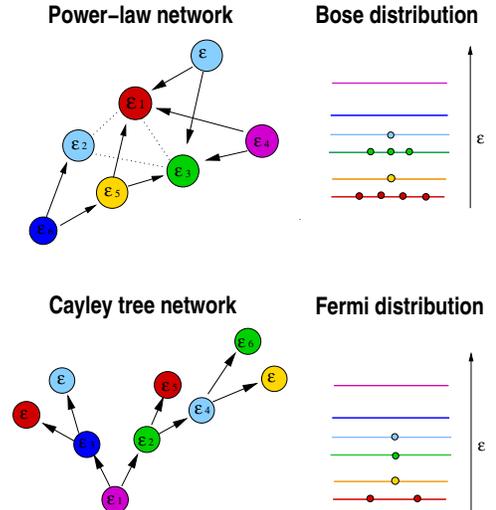}}
\caption{Symmetric construction of a power-law network and the
Cayley tree model considered in this paper. } \label{Treefig.fig}
\end{figure}

\section {The Model}

The Cayley tree  (or Bethe lattice)
 is a loop-free network in which there are three classes of nodes:
 {\it the root node}, which
 is at the origin of the tree and has connectivity $m$,
 the {\it nodes at the interface} with connectivity  one and
 the {\it nodes in the bulk } (below the interface) with connectivity $m+1$.

We start from the root of the tree (node $i=1$) and we link it to
$m$  new nodes $i=2,3, \dots ,m+1 $. We indicate each node with a
subsequent number,
  $t_i$ indicating the time in which it is arrived in the interface.

At each timestep  we choose  one  node to grow, giving rise to $m$
new nodes. Consequently, the interface of the tree  grows linearly
in time, and the growing node is chosen at each time from the
growing number of active ones. In order to mimic the quenched
noise of the medium we assign to each node of the tree  an energy
$\epsilon$ from a fixed random distribution $p(\epsilon)$.

\begin{figure}
\centerline{\epsfxsize=3.0in \epsfbox{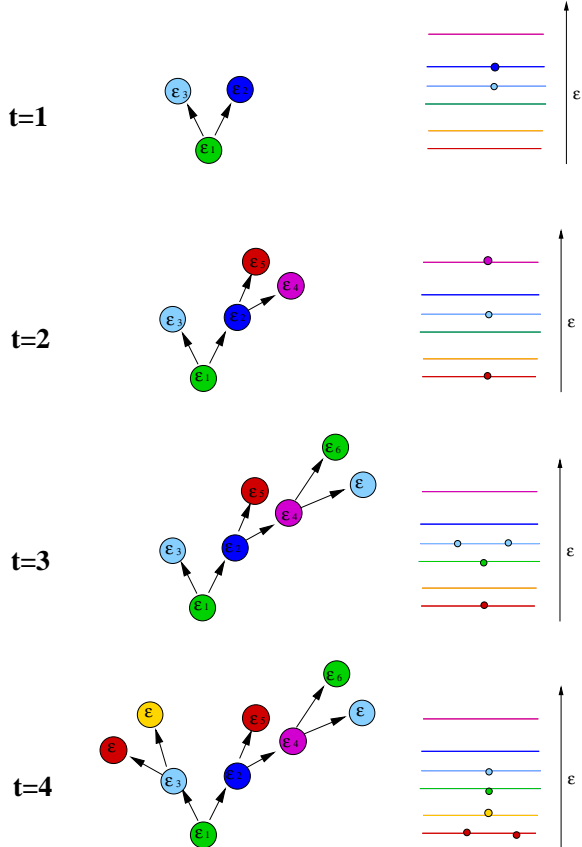}}
\caption{Description of the dynamics of the model $m=2$. At time
$t=1$ the root node $i=1$ with energy $\epsilon_1$ grows giving
rise to $m$ new nodes $i=2,3$ with energies $\epsilon_2,
\epsilon_3$. Node $i=1$ is in the bulk or below the interface
(connectivity $m$) while the nodes $i=2,3$ have connectivity one
and are at the interface. At time $t=2$ the node $i=2$ is chosen
to grow, it leaves the interface giving rise to $m$ new nodes.
$i=4,5$. At time $t=3,4$ nodes $i=4$ and $i=3$ are chosen to grow.
In the right hand side of the figure we plot the density of states
of the node in the interface. For example at time $t=4$ the nodes
at the interface are $i=5,6,7,8,9$ and we can draw the density of
states of the node at the interface by placing a particle in the
energies levels $\epsilon_5=\epsilon_8,
\epsilon_9<\epsilon_6<\epsilon_7$ as indicated in the picture.}
\label{TreeDyn.fig}
\end{figure}

We assume that  higher energy nodes are more likely to grow than
lower energy ones and that the probability $\Pi_i$ for the active
node $i$ (with energy $\epsilon_i$) to grow at time $t$ is given
by
\be
\Pi_i=\frac{e^{\beta \epsilon_i}}{\sum_{j\in Int(t)} e^{\beta
\epsilon_j}}. \label{TreeEn.pi1} \ee where the sum  in the
denominator is  extended to all  nodes $j$ that belong to the
interface $Int(t)$ at time $t$. The model  depends on the
parameter $\beta$. Tuning $\beta$ we change the nature of the
model and the spatial aspect of the
 tree. In the $\beta\rightarrow 0$ limit, high and low energy
 nodes are  equally probable to grow and the model reduces to the {\it Eden model}
  while in the $\beta\rightarrow \infty$ limit
the dynamics becomes extremal such that only the nodes with the
highest energy value are allowed to grow and the model reduces to
{\it invasion percolation }\cite{IP} on a Cayley tree.

\section{ Eden model on a Cayley tree}

 Let us assume that
 every node has the same energy $\epsilon^0$, i. e.
 $p(\epsilon)=\delta(\epsilon-\epsilon^0)$.
 In this case all nodes at the interface are equally likely to grow and we
call this model the Eden model on a Cayley tree. The probability
that a node $i$ of the interface $Int(t)$
 grows at time $t$ is given by
\be
\Pi_{i}=\frac{1}{N_{Int}(t)}, \label{TreeEden.pi}\ee where
$N_{Int}(t)$ is the total number of active nodes. Since at each
timestep a node of the interface grows, becoming part of the bulk,
and $m$ new active nodes  are generated, after $t$ timesteps the
model generates an interface of $N_{Int}(t)$ nodes, with
 \be
N_{Int}(t)=(m-1)t+1. \label{Interface.eq} \ee

 We denote by $\rho(t,t_i)$ the probability that a node,
 born at time $t_i$ is still active at time $t$.
Since every node grows with probability $\Pi_i$ Eq.
$\ref{TreeEden.pi}$ only if $i$ is a node of the interface,in mean
field $\rho(t,t_i)$ follows
\be
\frac{\partial \rho(t,t_i)}{\partial t}= -\frac{ \rho(t,t_i)}
{N_{Int}(t)}. \label{TreeEden.eq}\ee Substituting
$(\ref{Interface.eq})$ in $(\ref{TreeEden.eq})$  in the limit
$t\rightarrow \infty$ we get the solution
\be
\rho(t,t_i)={\left(\frac{t_i}{t}\right)}^{1/(m-1)}.
\label{TreeEden.dyn} \ee Consequently each node $i$ that  arrives
at the surface at time $t_i$, remains at the surface with a
probability that decreases  in time as a power-law. On the other
side the same power-law describes also
 the distribution of the age $\tau$
of the nodes at the interface observed at time $t$. In fact, the
probability $P(\tau)$ that a node   born at time  $\tau$ is still
active at time $t$, is given by
\be
P(\tau)={\left(\frac{\tau}{t}\right)}^{1/(m-1)}.
\label{TreeEden.tau} \ee Thus asymptotically in time the same
power-law
  describes the time  evolution
of the nodes  born at time $t_i$, ($\rho(t,t_i)$) and
 the age distribution of the nodes in the interface, ($P(\tau)$).

\subsubsection{Numerical simulations}
In order to verify the theoretical predictions, we have performed
numerical simulations of the Eden model on   a Cayley tree
 with
$m=2,4,6$. In Fig.~$\ref{Edentau.fig}$, we report the age
distribution $P(\tau)$ of the nodes at the interface for Cayley
trees with $m \times 10^4$ nodes and $m=2,4,6$. The data, averaged
over $100$ runs, follows the power-law predicted by
$(\ref{TreeEden.tau})$.  Numerical data are reported together with
the theoretically predicted power-law $(\ref{TreeEden.tau})$.

\begin{figure}
\centerline{\epsfxsize=3.5in \epsfbox{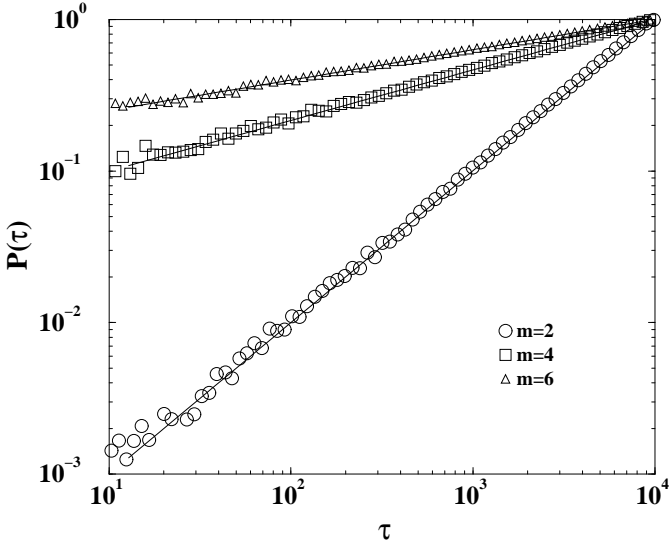}} \caption{Age
distribution $P(\tau)$ of the  nodes at the interface  of a Cayley
trees  with connectivities $m=2,4,6$ and $10^4$ generations. Data
have been averaged over $100$ runs. The solid lines represent the
power-laws predicted by $(\ref{TreeEden.tau})$ with exponent
$1-1/m$.} \label{Edentau.fig}
\end{figure}

\section{ Cayley tree with energies}

At  finite temperature, $\beta\neq 0$ it is necessary to take into
account the fact that each node has a different energy that
defines its dynamics. In this case the probability $\Pi_i$ for a
node  $i$ at the interface with energy $\epsilon_i$ to grow at
time $t$ is given by $(\ref{TreeEn.pi1})$
\be
\Pi_i=\frac{e^{\beta \epsilon_i}}{\sum_{j\in Int(t)} e^{\beta
\epsilon_j}}. \label{TreeEn.pi3} \ee

 Since only  nodes at the interface can  grow,
the probability that  node $i$ would leave the interface at time
$t$ is given by the product of $\rho_i(t|\epsilon_i,t_i)$ (the
probability that the node is active at time $t$) and $\Pi_i$ (the
probability that the node
 is chosen to grow in between the active nodes).
Consequently, $\rho_i(t|\epsilon_i,t_i)$ decreases in time
following
 \be \frac{\partial
\rho(t|\epsilon_{t_i},t_i)}{\partial t}= - \frac{e^{ \beta
\epsilon_{t_i}} \rho(t|\epsilon_{t_i},t_i)} {\sum_{j\in Int(t)}
e^{ \beta \epsilon_{j}} }, \label{TreeEn.eq} \ee were the sum in
the denominator is extended to all the nodes $j$ that are part of
the interface $Int(t)$ at time $t$.

\subsubsection{ Analytic solution}

To solve ($\ref{TreeEn.eq}$) we assume that  in the thermodynamic
limit the sum $Z^S$ in the denominator of the l.h.s. of Eq.
$(\ref{TreeEn.eq})$, given by
\be
Z^S=\sum_{j\in Int(t) } e^{ \beta\epsilon_j} \ee self averages and
converges to its mean value
\be
Z^S\rightarrow <Z^S>=\sum_{j=1,N}  e^{ \beta\epsilon_j}
\rho_j(t|\epsilon_{j},t_j). \ee Moreover, since $Z^S$ is an
extensive quantity we can self-consistently assume
\be
<Z^S>\stackrel{t\rightarrow\infty}{\rightarrow} z_F
t(1+O(t^{-\alpha})). \label{TreeEn.sum2} \ee Using
($\ref{TreeEn.sum2}$), the dynamic equation ($\ref{TreeEn.eq}$)
can be written as
\be
\frac{\partial \rho(t|\epsilon_{t_i},t_i)}{\partial t}= -
\frac{e^{ \beta \epsilon_{i}}}{z_F}
\frac{\rho(t|\epsilon_{i},t_i)}{t}. \label{TreeEn.eq2} \ee

Consequently we found that the  time evolution of
$\rho_i(t|\epsilon_i,t_i)$ follows a power-law,
\be
\rho_i(t|{\epsilon_i},t_i)= \left( \frac{t_i}{t}
\right)^{f(\epsilon_i)}, \label{TreeEn.dyn} \ee
 but there is
multiscaling in the system, i.e. the dynamic exponent depends on
the energy $\epsilon_{i}$ of the node,
\be
f(\epsilon)=\frac{e^{\beta\epsilon}}{z_F}. \label{TreeEn.chi} \ee

After substituting $\rho_i(t|\epsilon_i,t_i)$ from Eq.
$(\ref{TreeEn.dyn})$ with $f(\epsilon)$ given by Eq.
($\ref{TreeEn.chi}$), into Eq. ($\ref{TreeEn.sum2}$), and the sum
with an integral, we get the self-consistent equation

\bea
 <\sum_j e^{\beta\epsilon_j} \rho_j(t|\epsilon_j,t_j)>&\stackrel{t\rightarrow\infty}{=}&
 \int d\epsilon \ \ p(\epsilon) \int_1^t dt' e^{\beta \epsilon} \left( \frac{t'}{t}
\right)^{f(\epsilon)}
 \nonumber \\&=&  z_F  t
 (1+O(t^{-\alpha}),
\label{TreeEn.sum} \eea where \bea
\alpha=(1+\min_{\epsilon}f(\epsilon))>1,
 \\
 z_F=m \int {d\epsilon p(\epsilon) \frac{
e^{\beta\epsilon}}{1+f(\epsilon)}} . \label{TreeEn.z} \eea
Finally, if we define $\mu_F$ as
\be
z_F=e^{\beta \mu_F}, \ee we obtain the self-consistent Eq.
($\ref{TreeEn.z}$) can be interpreted as a definition of $\mu_F$
and it is formally equivalent to the definition of the chemical
potential in an {\it equilibrium Fermi gas}
\be
1-\frac{1}{m}= \int d\epsilon
p(\epsilon)\frac{1}{e^{\beta(\epsilon-\mu_F)}+1},\label{TreeEn.selfc}
\ee suggesting  that many properties of this model can be
described by the Fermi statistics.

On the other hand the probability $P(\tau)$ that a node   born at
time $\tau$ is still active at time $t$, is given by a power-law
\be P(\tau)=\int d\epsilon
p(\epsilon){\left(\frac{\tau}{t}\right)}^{e^{\beta(\epsilon-\mu_F)}}\sim
{\left(\frac{\tau}{t}\right)}^{\delta}. \label{TreeEn.tau} \ee

\subsubsection{ Existence of the solution for the chemical potential}

Equation ($\ref{TreeEn.selfc}$)  always has a solution for the
chemical potential $\mu_F$. In fact, since $p(\epsilon)$ is a
normalized distribution function,
 the integral
 $I\{ p(\epsilon),\mu_F\}$ on the r.h.s. of equation
($\ref{TreeEn.selfc}$),
\be
I\{ p(\epsilon),\mu_F\}=\int d\epsilon p(\epsilon)n_F(\epsilon)
\ee
 for a fixed  $\mu_F$ is bounded by
 \be
\frac{1}{e^{\beta (\epsilon_{min}-\mu_F)}+1}<I\{p(\epsilon),
\mu_F\}<\frac{1}{e^{\beta(\epsilon_{max} -\mu_F)}+1} \ee implying
that for the chemical potential $\mu_F$ the solution of
Eq.($\ref{TreeEn.selfc}$), $I\{p(\epsilon), \mu_F\}=1-1/m$,
satisfies
\be
\epsilon_{min}+\frac{1}{\beta}\log(m-1)<{ \mu_F}<
\epsilon_{max}+\frac{1}{\beta}\log(m-1). \ee This proves  that for
a real tree with $m>1$ the
  equation ($\ref{TreeEn.selfc}$)
  always has a solution.
Finally some attention should be given to the special limits
$\beta\rightarrow 0$ and $\beta\rightarrow\infty$.

{\it $\beta\rightarrow 0$ limit --} In this case we recover the
solution of the Eden model on the tree, $z_F=m-1$. Since the
probability distribution $p(\epsilon)$ is normalizable and the
occupation number
\be
n_F(\epsilon)\rightarrow \frac{1}{z_F^{-1}+1},
\label{TreeEn.ncost} \ee equation  ($\ref{TreeEn.selfc}$) reduces
to $z_F\rightarrow m-1$ and thus
 $\beta \mu_F\rightarrow \ln(m-1)$
in such a way that  $\mu_F\ge 0$ ($z_F>1$) if $m>2$.

{\it  $\beta \rightarrow \infty$ limit--}
 In this limit the Fermi-Dirac distribution converges to the
step function
\be
n_F(\epsilon)\rightarrow \theta(\epsilon-\mu_F)
\label{TreeEb.ntheta} \ee and the self-consistent equation
($\ref{TreeEn.selfc}$) becomes
\be
1-\frac{1}{m}=\int_{\epsilon<\mu_F} p(\epsilon).
\label{TreeEn.binf}\ee

\subsubsection{ Mass conservation}

The self-consistent relation $(\ref{TreeEn.selfc})$ can also be
derived  from mass conservation, i.e. from the knowledge that
 the total number of nodes at the interface is given by $N=(m-1)t$.
Consequently,
\be
N=(m-1)t=\sum_{i\in[0,mt]} \rho(t|\epsilon_i,t_i).
\label{TreeEn.massc} \ee We can substitute the sum in the right
hand side of Eq.~($\ref{TreeEn.massc}$) with the mean over the
energies  $\epsilon_i$ of the nodes $i$ of generation $t_i$.
Moreover, in the thermodynamic limit we can approximate the sum
over $i$ with an integral over $t_i$, the mass conservation
relation becoming
 \bea (m-1)t&=&m\int d\epsilon p(\epsilon)
\int_1^{t} d\tau
 \rho(t|\epsilon, \tau)
\nonumber\\ &=& m \int d\epsilon p(\epsilon) \int_1^t d\tau \
 {\left(\frac{\tau}{t}\right)}^{e^{\beta(\epsilon-\mu_F)}} \nonumber \\
& \simeq &m t \int d\epsilon
p(\epsilon)\frac{1}{e^{\beta(\epsilon-\mu_F)}+1}
\label{TreeEn.masscf} \eea where in the last equation we have
neglected terms of order $O(t^{-\alpha})$. Thus both the mass
conservation relation ($\ref{TreeEn.massc}$) and the
self-consistent relation ($\ref{TreeEn.selfc}$)  allow us to
define the chemical potential $\mu_F$, describing the evolution of
the network as the chemical potential of an equilibrium Fermi gas
with specific volume $v_c=1+1/(m-1)$. However this last expression
explains the meaning of that relation. In fact, the number
$N_{Int}(\epsilon)$ of nodes with energy $\epsilon$ at the
interface at  time $t$
 is given by
\be
N_{Int}(\epsilon)=mt n_F(\epsilon)p(\epsilon), \label{TreeEn.nit}
\ee where $n_F(\epsilon)$ is given by the Fermi occupation number
\be
n_F(\epsilon)=\frac{1}{e^{\beta(\epsilon-\mu_F)}+1}.
\label{TreeEn.nf} \ee In other words, the distribution of the
energy at the interface reaches a stationary limit given by
$(\ref{TreeEn.nit})$ and defined by a Fermi distribution with
chemical potential given by ($\ref{TreeEn.selfc}$). In the mean
time,  the density of nodes with energy $\epsilon$ present in the
bulk, $N_{Bulk}(\epsilon)$, reaches a stationary limit as well. In
fact, since the nodes in the bulk are the ones of the network that
are not at the interface, using $(\ref{TreeEn.nit})$, we have
\be
N_{Bulk}(\epsilon)=p(\epsilon)[1-n_F(\epsilon)].
\label{TreeEn.nbulk} \ee

\subsubsection{ Asymptotic dynamics}

The dynamical evolution of the network brings the system to the
stationary state\cite{FST,RTS} described by the distribution
function ($\ref{TreeEn.nf}$), as it has been shown by the solution
of the dynamical equation $(\ref{TreeEn.eq})$. Moreover the
dynamics stabilizes this distribution. In fact, in the asymptotic
limit, when the survivability follows   ($\ref{TreeEn.dyn}$) the
probability that a node of energy $\epsilon$ will grow and leave
the interface is given by
\be
\pi_F(\epsilon,t)=m \int d\epsilon'  p(\epsilon')\int_1^t dt'
\frac{\partial \rho(t|\epsilon',t')}{\partial t}
\delta(\epsilon-\epsilon') \label{TreeEn.pidef} \ee which, using
($\ref{TreeEn.eq2}$) can be extimated to be \bea
\pi_F(\epsilon,t)&=&\int d\epsilon' m p(\epsilon') \int_1^t dt'
\frac{e^{\beta(\epsilon-\mu_F)} \rho(t|\epsilon',t')}{t}
\delta(\epsilon-\epsilon') \nonumber \\ &=& m
e^{\beta(\epsilon-\mu_F)}  p(\epsilon)\frac{1}{t}\int_1^t dt'
  {\left(\frac{t'}{t}\right)}^{e^{\beta(\epsilon-\mu_F)}} \nonumber \\
&\simeq & m p(\epsilon) [1-n_F(\epsilon)] .\label{TreeEn.pias}
\eea Consequently,  the probability that a node of energy
$\epsilon$ leaves the interface, asymptotically in time, reaches a
stationary limit
 independent of the particular evolution of the network, given by
\be
\pi_F(\epsilon,t) \rightarrow \pi_F^*(\epsilon)
=p(\epsilon)[1-n_F(\epsilon)]. \label{TreeEn.pias2} \ee If we
observe an
 evolving network and we have no knowledge of the
 age of the nodes, but only of their energies,
 the complete dynamics  is determined by  $\pi(\epsilon,t)$ describing
which is the probability that a node with energy $\epsilon$ will
leave the interface at  time $t$. While the complete dynamics
$(\ref{TreeEn.eq})$ is clearly dependent on time,
$\pi(\epsilon,t)$,  reaches the stationary limit
$\pi_F^*(\epsilon)$, defining the invariant dynamics of the
system.

The stability of the distribution  $N_{bulk}(\epsilon)$ of the
energies in the bulk, is thus enforced by the dynamics. In fact we
have found that asymptotically in time the probability that a node
with energy $\epsilon$ is chosen to grow  $\pi_F^*(\epsilon)$ is
proportional to the number of nodes in the bulk
$N_{bulk}(\epsilon)$ given by Eq. $(\ref{TreeEn.nbulk})$.

\subsubsection{Numerical support}

Choosing  the node energy from a uniform distribution
$p(\epsilon)=1$ with $\epsilon\in [0,1]$, we have  simulated the
growth of a Cayley tree with  $m=2$ and various values of $\beta$.
In Fig.~$\ref{Treefermi.fig}$ we report the distribution of the
energies of the active nodes for a network of size $N=2\times
10^4$ nodes for $\beta=5,10,30$. The solid line in the figure
represents the theoretical prediction described by
$(\ref{TreeEn.nit})$ and $(\ref{TreeEn.nf})$ with a chemical
potential given by $(\ref{TreeEn.selfc})$. In
Fig.~$\ref{per-tau.fig}$ the distribution of the age of the nodes
at the interface  is shown for  $\beta=2,5,10,20$ and compared to
the theoretical prediction Eq.$(\ref{TreeEn.tau})$,
$P(\tau)\sim\left(\frac{\tau}{t}\right)^{\delta}$ with $\delta$
given by $e^{-\beta \mu_F}$ for the uniform distribution
$p(\epsilon)=1$, $\epsilon\in [0,1]$.

\begin{figure}
\centerline{\epsfxsize=3.5in \epsfbox{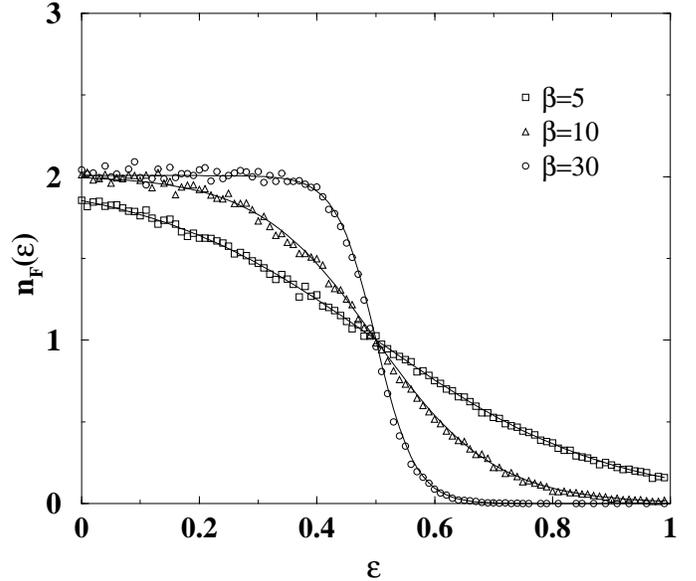}} \caption{The
energy distribution of the nodes in the interface for
$\beta=5,10,30$ in the case of a uniform energy distribution
$p(\epsilon)=1$ for $\epsilon \in [0,1]$, $m=2$ and predicted
chemical potential $\mu_F=1/2$.The solid lines indicate the
predicted Fermi distribution.} \label{Treefermi.fig}
\end{figure}

\begin{figure}
\centerline{\epsfxsize=3.5in \epsfbox{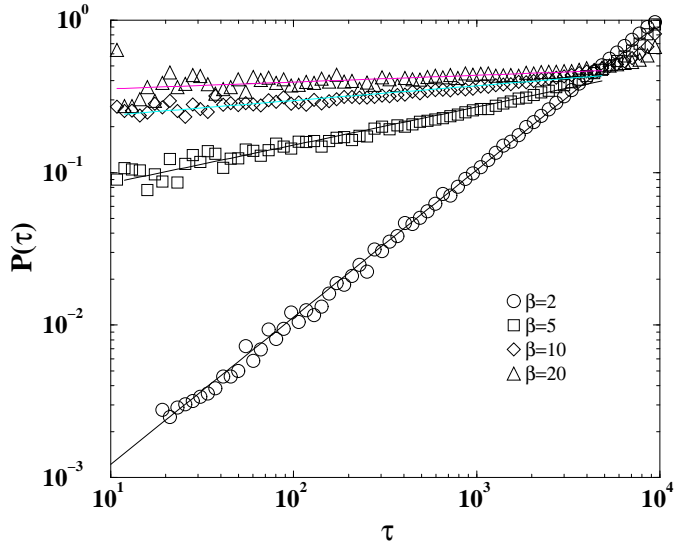}}
\caption{Probability distribution of the age $\tau$ of the nodes
at the interface in a Cayley tree with $m=2$ and time $10^4$ as a
function of $\beta=2,5,10,20$.} \label{per-tau.fig}
\end{figure}

\section{Connection with other problems in statistical mechanics}

\subsection{Invasion Percolation}

 The static percolation problem on a Cayley tree
is exactly solved and gives for the percolation threshold
$p_c=1-1/m$. The invasion percolation problem, represents the
first example of a self-organized system and it has been solved
exactly on a Cayley tree in \cite{IPCT} and numerically in
\cite{Vanderwalle}.

If we suppose  that the energies of the tree model are uniformly
distributed between zero and one, i.e. $p(\epsilon)=1$ for
$\epsilon\in [0,1]$, in the limit $\beta=\infty$ the evolution of
the interface of the Cayley tree network maps exactly into  the
front of  invasion percolation  on a Cayley tree. We can find the
chemical potential $\mu_F$ by solving  \bea 1-1/m&=&\int d\epsilon
n_F(\epsilon) \nonumber \\ &=& \int_0^1 d\epsilon
\frac{1}{e^{\beta (\epsilon-\mu_F)}+1}\nonumber \\
&=&\int_0^{\mu_F} d\epsilon = \mu_F. \label{IP}\eea Consequently,
 asymptotically in time, the density of nodes
 with energy $\epsilon$ found at the interface
follows the  step function
\be
n_F(\epsilon)=\theta(\epsilon-(1-1/m)), \label{TreeIP.nsf}\ee
predicting the correct threshold $p_c=1-1/m$ for  invasion
percolation on a Cayley tree.

\subsection{Percolation}

The percolation transition can be investigated choosing a
distribution function
\be
p(\epsilon)=p\delta(\epsilon-1)+(1-p)\delta(\epsilon),
\label{Treeper.pe} \ee describing the fact that each node of the
Cayley tree has energy $\epsilon=1$ with probability $p$ and
energy $\epsilon=0$ with probability $1-p$. In the limit
$T\rightarrow 0$ only the nodes with $\epsilon=1$ would grow.
Consequently, assigning to the energy  at the interface a meaning
of cost function $\cite{CT}$ we have that in the $T=0$ the mean
energy of the node at the interface is minimized. When we consider
the probability distribution $(\ref{Treeper.pe})$, the
self-consistent equation $(\ref{TreeEn.selfc})$ reduces to a
quadratic equation for the fugacity $z_F$,
\be
(1-\frac{1}{m})=\frac{1-p}{z_F^{-1}+1}+\frac{p}{e^{\beta}z_F^{-1}+1}.
\ee that in the  $\beta \rightarrow \infty$ limit has the
solutions \bea
z_F^{-1}&=&\frac{-\Delta}{2(1-p_c)}\left(1\pm\sqrt{1+4 e^{-\beta}
\frac{p_c(1-p_c)}{\Delta^2}}\right)\label{solutions} \eea with
$\Delta=(p-p_c)$ and $p_c=1/m$.  But only one  solution is
possible, because the fugacity is strictly positive. This implies
that if $\Delta$ is positive ($p>p_c$) we should take the negative
sign in equation $(\ref{solutions})$ while we have to use the
positive sign for $p<p_c$.

Thus we have,
\be
z_F e^{-\beta}\simeq\left\{\begin{array}{lr}
e^{-\beta}(1-p_c)/{|\Delta|} & p<p_c \\
 |\Delta|/p_c & p>p_c\end{array}\right..
\label{Tree.perz} \ee  Consequently, using ($\ref{Treeper.pe}$),
the mean energy of the nodes at the interface is: \bea
\epsilon_{mean}&=&\frac{1}{m-1}\int d\epsilon \ \epsilon \
p(\epsilon)n_F(\epsilon)\nonumber\\
&=&\frac{p_c}{1-p_c}\frac{p}{e^{\beta}z_F^{-1}+1}.
\label{Tree.emean} \eea In the $T\rightarrow 0$ limit
$\epsilon_{mean}$ displays a sharp transition at $p_c=1/m$. Using
the solution $(\ref{Tree.perz})$ in $(\ref{Tree.emean})$, we
obtain
\be
\epsilon_{mean}=\left\{\begin{array}{lr} e^{-\beta}
{pp_c}/{|\Delta|}  & p<p_c \\ \Delta/(1-p_c) &
p>p_c\end{array}\right. \ee
\begin{figure}
\centerline{\epsfxsize=3.5in \epsfbox{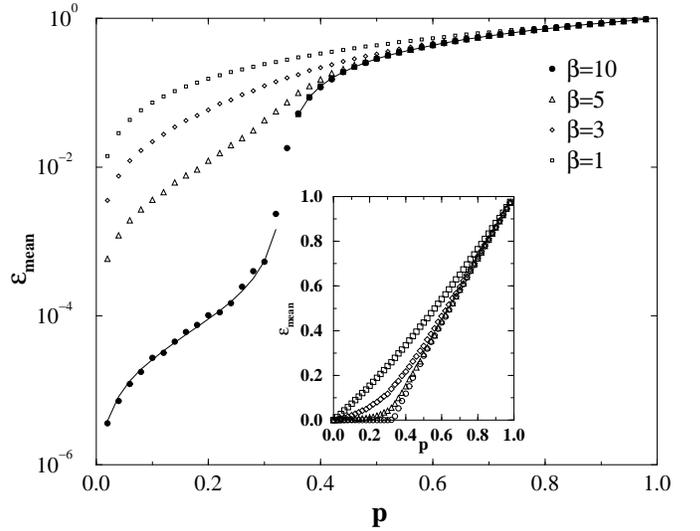}} \caption{Mean
energy of the nodes at the interface  as a function of $p$ and
$\beta=10,5,3,1$. The solid line represent the theoretical
prediction in the $\beta\rightarrow \infty$ limit.}
\label{TreePer.fig2}
\end{figure}
In order to  verify the predicted solution  we have simulated the
growth of a Cayley tree with $m=3$ branches.  We have assigned to
each node an energy equal to $0$ or $1$
 with probability
given by $(\ref{Treeper.pe})$.  The evolution of the network
described by ($\ref{TreeEn.pi1}$) proceeds preferentially on high
energy nodes, in such a way that in the $\beta \rightarrow \infty
$ the mean energy of the interface is minimized. In
Fig.~$\ref{TreePer.fig2}$ we report the mean energy of the nodes
at the interface  as a function
 of $p$ for different temperatures.
As the temperature goes to zero, the mean energy of the system has
a sharp transition from zero to a non zero mean energy
distribution. This shows that during the  network  evolution the
nodes with high energy  remain at the interface.

\section{Partition in time, energy and measure}

\subsection{Partition in time}

If we coarse grain the time and divide the nodes following their
age, we observe that the nodes with approximately the same age
have a survivability $\rho(t,t_i)$ that grows in time as a
power-law of $\ell_i=t/t_i$,
\be
<\rho(t,t_i)> _{\epsilon}=\int d\epsilon p(\epsilon)
\rho(t,t_i,\epsilon)\simeq{\ell_i}^{-\tau_F}
\label{Con.pl.time}\ee with  $\tau_F$ asymptotically given by
\be
\tau_F \simeq e^{\beta \mu_F}. \ee This power-law describes the
fact  that the probability
 to be at the interface is lower for older nodes, but decreases
much faster for nodes of energy $\epsilon>\mu_F$. The scale of the
power-law depends on the ratio between the observation time and
the time $t_i$  when the node $i$ joins the interface:
\be
\ell_i=\frac{t}{t_i}. \label{Con.ella} \ee Because of this
structure $\rho(t,t_i)$ is not a function separately of  $t$ and
$t_i$, and not even of their difference $t-t_i$, but only of their
ratio $\ell_i$. Thus $(\ref{Con.pl.time})$ describes both the
probability that a node born at time $t_i$ is at the interface at
time $t$  and the probability that a node at the interface at time
$t$ is arrived in the network at time $t_i$.

Consequently  the network at a given time will contain nodes of
every age at different stages of their evolution. In other words
the tree acts as if it was recording the evolution of the nodes,
and the time partitions of the Cayley tree behaves as  riscaled
copies of the same network.

\subsection{Partition in energy}

A different case implies that the ages of the
 nodes are unknown but we have information about their energies.Thus
 it is natural to  group together nodes with the same energy.

Since the total number of nodes with energy $\epsilon$ in the
Cayley tree
 are  $p(\epsilon)t$, the fraction of the nodes that remain at the interface
is given by the integral of  $\rho(t|t_i,\epsilon)$ over $t_i$,
normalized
 to $p(\epsilon)t$, i.e.
\be
\frac{1}{p(\epsilon)t}\int dt_i p(\epsilon)
\rho(t|t_i,\epsilon)=n_F(\epsilon). \ee As  observed  above this
density is given by the Fermi occupation number $n_F(\epsilon)$,
reached in the thermodynamic limit, and it gives   the
characteristic stable distribution  of the Cayley tree evolution.

\subsection{The partition functions }

The emergence of the quantum statistics in the description of the
geometrical structure of the  interface of the Cayley
 tree suggests us
the possibility to perform a more detailed investigation of its
statistical properties.

\subsubsection{Static partition function}

In order to study the aspect of the Cayley tree at a given time,
we
 can define the static partition function $Z^S(t)$ as
\be
Z^S(t)=\frac{\sum_{j\in Int(t)} e^{\beta \epsilon_j} }{(m-1)}
\label{Con.zs} \ee This  describes the distribution of the energy
values in the interface. As proven previously in
$(\ref{TreeEn.z})$, the static partition function, asymptotically
in time, reaches a stationary limit and satisfies
\be
\lim_{t\rightarrow \infty} \frac{Z^S(t)}{t}=\frac{z_F}{(m-1)} \ee
Consequently, the static partition function  defined in
Eq.~$(\ref{Con.zs})$ is an extensive quantity related to the
chemical potential of the network by
\be
\frac{1}{\beta}\lim_{t\rightarrow \infty}
\frac{\log(Z^S(t))}{t}=\mu_F-T\log(m-1). \ee

\subsubsection{Dynamic partition function}

In order to describe the Cayley tree evolution  we associate  to
it the sequence $\{\epsilon^F_r\}$ of the energy of the nodes
selected to grow at time $t=r$. If we consider all the
 time dependent sequences
associated to different realization of the networks  for $t$
timesteps, we can associate with them a dynamical partition
function $Z^D(t)$ given by
\be
Z^D(t)=\int\Pi_{r=1,t}\{d\epsilon_r\} e^{\beta
\sum_{r=1}^{t}\epsilon_r}
P(\epsilon_1,\epsilon_2,\dots\epsilon_{t}) \label {Const.zdin} \ee
where the sum goes over all the different realizations of the
network and $P(\epsilon_1,\epsilon_2,\epsilon_{mt})$ is the
probability of the sequence $\{\epsilon_r\}$. As we have shown
previously (Eq.~$\ref{TreeEn.pias2}$), in the thermodynamic limit
the probability that an energy $\epsilon$
 will be selected to grow reaches a stationary value given by the quantum occupation numbers.
Consequently the partition function, asymptotically in time can be
written as
\be
Z^D_F(t+1)\simeq  Z^D_F(t)\int d\epsilon e^{\beta
\epsilon}\pi^*_{F}(\epsilon) \ee
 giving in the thermodynamic limit,
\be
Z^D(t)\simeq (z_F^D)^t \ee were
\be
z_F^D=m\int d\epsilon p(\epsilon)e^{\beta
\epsilon}[1-n_F(\epsilon)]. \ee Thus for the induced dynamic in
the energy space we can introduce a dynamical partition function
describing the possible realization of the networks. The dynamical
partition function,  asymptotically in time is given by
\be
\lim_{t\rightarrow \infty}\frac{\log(Z^D_F(t))}{t}=\log(z_F^D) \ee

The connection between the dynamical and the static fugacity is
easily shown to be,
\be
(m-1)z_F^S+z_F^D=m<e^{\beta \epsilon}>_{p(\epsilon)} \ee

\subsubsection{Entropy}

One characteristic  feature of this model is that the dynamics
introduces a distinction between  different nodes
 of the tree.
In fact the nodes of the tree are distinguished  between nodes in
the interface and nodes in the bulk. The separation of this 'two
phases' is forced by the dynamics itself and can be the reason for
the statistical mechanics properties  described by quantum
statistics. In fact, if we look at the structure of the fermionic
entropy $S$ , \bea S=-\int d\epsilon p(\epsilon)n_F(\epsilon)
\log(n_F(\epsilon))+\nonumber
\\ -\int d\epsilon p(\epsilon)
[1-n_F(\epsilon)]\log([1-n_F(\epsilon)]) \eea we can observe that
it can be interpreted as the  sum of two Shannon entropies of the
tree: the bulk and of the interface entropy,
\be
S_F=S_{int}+S_{bulk}. \ee

\section{Conclusions}

In this work we have introduced a model for a growing Cayley tree
with thermal noise characterized by
\begin{itemize}
\item
{\it Growth:}   At each time exactly $m$ nodes are added  and one
is eliminated at the interface, the number of nodes in which
percolation can occur  grows in linearly in time as $N=(m-1)t$
nodes;
\item
{\it Time dependent dynamics:} Each  node can percolate only once
and the probability for a node to be chosen as the percolating one
is a decreasing function of  time.
\end{itemize}
 We have solved analytically the model studying
its character in particular on the limit $\beta=0$ and then at
finite temperature. The distribution of strength bonds  follows a
Fermi distribution  and the dynamics asymptotically in time
replicates and stabilizes this distribution. The bond strength
plays the role of energies in the Fermi distribution. The
distribution of ages of the node at the interface follows an
effective power-law. We have then studied the cases in which the
model reduces to the study of Invasion Percolation and
percolation. We have found the percolation threshold and the mean
energy of the growth. Finally we investigate the statistical
properties of the solution and we define two partition functions
describing the ensemble of growing Cayley trees. Finally this
system is a symmetric construction of a power-law network
following Bose distribution, as shown in Fig.1 and it opens the
way to understand the self-organized nature of scale-free
networks. \cite{sym}.

\section{Akwnoledgements}

I would like to thank prof. A.-L. Barab\'asi for  support  and
help and L. Pietronero and A. Gabrielli for useful discussions.


\end{multicols}
\end{document}